\documentclass[preprint,showpacs,preprintnumbers,amsmath,amssymb,superscriptaddress]{revtex4}

\usepackage{graphicx}% Include figure files
\usepackage{dcolumn}% Align table columns on decimal point
\usepackage{bm}% bold math

\newtheorem{theorem}{Theorem}
\newenvironment{proof}[1][Proof]{\begin{trivlist}
\item[\hskip \labelsep {\bfseries #1}]}{\end{trivlist}}
\newcommand{\qed}{\nobreak \ifvmode \relax \else
      \ifdim\lastskip<1.5em \hskip-\lastskip
      \hskip1.5em plus0em minus0.5em \fi \nobreak
      \vrule height0.75em width0.5em depth0.25em\fi}

\begin{document}

\preprint{q-bio.NC/0404021}

\title{Pulse-coupled resonate-and-fire models}

\author{Keiji Miura}
\email{kmiura@brain.riken.jp}
\affiliation{Department of Physics, Graduate School of Sciences, Kyoto University Kyoto 606-8502, Japan}
\affiliation{Laboratory for Mathematical Neuroscience, RIKEN Brain Science Institute, Saitama 351-0198, Japan}
\author{Masato Okada}
\affiliation{Laboratory for Mathematical Neuroscience, RIKEN Brain Science Institute, Saitama 351-0198, Japan}
\affiliation{``Intelligent Cooperation and Control'', PRESTO, JST,\\c/o RIKEN BSI, Saitama 351-0198, Japan}

\begin{abstract}
We analyze two pulse-coupled resonate-and-fire neurons.
Numerical simulation reveals that an anti-phase state
is an attractor of this model.
We can analytically explain the stability of anti-phase states
by means of a return map of firing times, which we propose in this paper.
The resultant stability condition turns out to be quite simple.
The phase diagram based on our theory shows that
there are two types of anti-phase states.
One of these cannot be seen in coupled integrate-and-fire models and
is peculiar to resonate-and-fire models.
The results of our theory coincide with those of numerical simulations.
\end{abstract}


\maketitle

\section{\label{sec1}Introduction}
The integrate-and-fire model \cite{koch} is well known in the context of 
spiking neuron models.
However, it cannot reproduce voltage oscillations near the equilibrium 
state and resonance in response to sinusoidal current inputs 
seen in the Hodgkin-Huxley model \cite{izhikevich,izhikevich2,izhikevich3}.
Although we need a model with more than two variables to reproduce these 
phenomena \cite{izhikevich4},
in general, it is difficult to solve models with more than two variables 
analytically.

Izhikevich suggested the resonate-and-fire model, which can reproduce
voltage oscillations and resonance and is still analytically tractable
\cite{izhikevich}.
The resonate-and-fire model is a 2-dimensional linear dynamical system with 
a threshold.

Voltage oscillations can play an important role in transmitting signals 
in the brain.
For example, Izhikevich pointed out that resonance provided an effective tool
in selective communication \cite{izhikevich,izhikevich2,izhikevich3}.
Therefore, we need to find out whether a network of resonate-and-fire 
models has properties that cannot be observed in 
integrate-and-fire models.
The network properties of resonate-and-fire-like models 
(linearized FitzHugh-Nagumo models) have been investigated \cite{garbo},
but only an oscillatory regime where individual neurons fire spontaneously 
was considered and subthreshold voltage oscillations were not focused on.

In this paper, we analyze a system of two 
excitatory or inhibitory pulse-coupled resonate-and-fire models.
We found that the system settled into an anti-phase state in both 
excitatory and inhibitory coupled cases in numerical simulations.
We theoretically evaluated the stability of anti-phase states in detail.
The system of pulse-coupled integrate-and-fire models has already been 
investigated \cite{strogatz}, 
and it has been proved that for almost all initial conditions the 
system evolves to an in-phase state.
This contrasts with resonate-and-fire models where the 
system does not necessarily evolve to an in-phase state.

In Sec.~\ref{sec2}, we briefly review the resonate-and-fire model 
and its properties.
We then suggest a system for two pulse-coupled resonate-and-fire models.

In Sec.~\ref{sec3}, we demonstrate the existence of an anti-phase state and 
construct a theory that can be used to determine the region of 
existence for anti-phase states and assess their stability.
We suggest an effective method of calculating the region.
We show that global stability of anti-phase states can be 
determined by return maps and their local stability can be determined 
by a simple equation.

In Sec.~\ref{sec4}, using the method proposed in Sec.~\ref{sec3}, 
we calculate a phase diagram for the existence and local stability of 
anti-phase states.
Stability switches near the boundary of the region of existence or 
at points where coupling strength is zero.
We also determine stability by direct numerical simulations, and 
the results coincide with those of our theory.
The phase diagram indicates there are two types of anti-phase states.
The anti-phase state with a longer period is unique to 
coupled resonate-and-fire models and 
cannot be found in coupled integrate-and-fire models.

\section{\label{sec2}Model}
\begin{figure}[t]
\includegraphics[width=70mm]{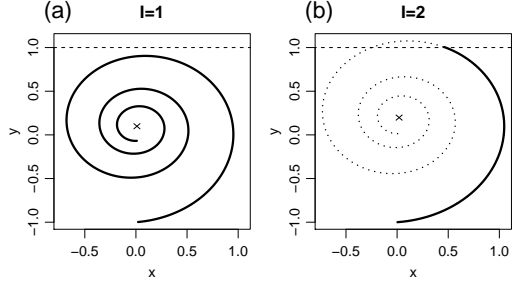}
\caption{\label{orbit.ps}
Solutions for resonate-and-fire model.
Orbits on xy plane are plotted.
Cross represents fixed point.
Dashed line is threshold.
(a) Orbit for $I=1$. Neuron does not fire.
(b) Orbit for $I=2$. Neuron fires. Neuron is immediately reset to 
$(x,y)=(0,-1)$ after it exceeds threshold.
Dotted line is orbit neuron follows if not reset.}
\end{figure}

A resonate-and-fire model is a 2-dimensional linear dynamical system 
with a threshold,
\begin{equation}
\left\{ \begin{array}{c}
\frac{dx}{dt}=-x-10y+I\\
\frac{dy}{dt}=10x-y,\end{array}\right.
\end{equation}
where $x$ and $y$ are internal state variables and $I$ is an external input.
If $y$ exceeds the threshold $(y=1)$, the internal state is reset to $(0,-1)$.
Fig.~\ref{orbit.ps} shows typical solutions.
The neuron with $I=1$ does not fire, while the neuron with $I=2$ exceeds the 
threshold and fires.
There is a critical value $I^1_C = 1.56$, and neurons with $I>I^1_C$ can fire.

The resonate-and-fire model has a fixed point satisfying,
\begin{equation}
\left\{ \begin{array}{c}
0=\frac{dx}{dt}=-x-10y+I\\
0=\frac{dy}{dt}=10x-y.\end{array}\right.\\
\end{equation}
The fixed point is $(x,y)=(\frac{I}{101},\frac{10I}{101})$.
When $I>I^2_C=10.1$, the fixed point is above the threshold and 
there is no stationary state.
The eigenvalues of a linearized system at the fixed point are 
$-1\pm10\mathrm{i}$.
Since the real part of the eigenvalue is negative, 
the fixed point is always stable if it exists.
The imaginary part represents the angular velocity around the fixed point.

In this paper, we analyze a system of two pulse-coupled resonate-and-fire 
models,
\begin{equation}
\left\{ \begin{array}{c}
\frac{dx_{i}}{dt}=-x_{i}-10y_{i}+I+K\sum\delta(t-t_{k})\\
\frac{dy_{i}}{dt}=10x_{i}-y_{i},\end{array}\right.(i=1,2)
\end{equation}
where $K$ is coupling strength.
A neuron receives a pulse in the $x$-direction 
at the moment another neuron exceeds the threshold $(y=1)$.
The neuron that has fired is immediately reset to $(x,y)=(0,-1)$.
Our reset value is different from the original value, $(x,y)=(0,1)$, 
in Izhikevich\cite{izhikevich}.
The reason is that with the original value, autocatalytic growth in the firing 
rate would accelerate indefinitely, leading to an explosion of the system.
To avoid this, we chose a reset value that did not lie directly on the 
threshold.
Since a neuron with our reset value has a refractory period,
our reset value is suitable for a neuron model.

\section{\label{sec3}Theory of anti-phase states}
\subsection{Anti-phase states}

\begin{figure}[t]
\includegraphics[width=70mm]{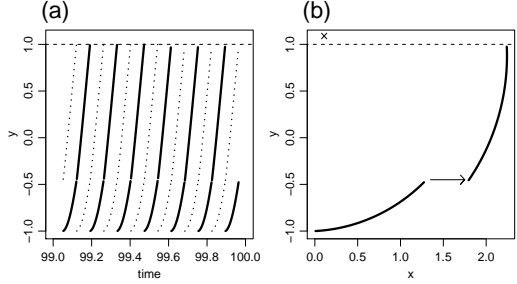}
\caption{\label{anti.ps}Example of anti-phase states. $K=0.5,I=11$. 
(a)Time evolution of y.
Horizontal axis represents time and vertical axis represents y.
Solid lines and dotted lines represent each neuron.
Dashed line is threshold $(y=1)$.
Two neurons fire alternately at regular intervals.
(b)Orbits in xy-plane.
Arrow indicates jump in $x$-direction by pulse input and 
cross denotes fixed point, $(x,y)=(\frac{11}{101},\frac{110}{101})$.
Dashed line is threshold line.
Here, both neurons follow same orbit, so two orbits overlap.}
\end{figure}

We simulated the system of coupled resonate-and-fire models 
with $K=0.5, I=11$, and randomized initial conditions 
and found that it evolves to the state shown in  Fig.~\ref{anti.ps}(a).
The solid lines and the dotted lines in the figure represent each neuron.
The dashed line is the threshold $(y=1)$.
The two neurons fire alternately at regular intervals.
Since each neuron fires periodically, a phase can be defined with
period $2\pi$.
We describe the firing time of one neuron as phase 0, which 
evolves in proportion to time.
Here, the time at which another neuron fires is 
described by phase $\pi$.
We call this state an anti-phase state.
In general, we refer to a state where two neurons follow the same orbit 
but have phase shift $\pi$ as an anti-phase state
\cite{pikovsky,kuramoto,winfree}.

Fig.~\ref{anti.ps}(b) plots the orbits in the $x$-$y$ plane for the same data 
as in Fig.~\ref{anti.ps}(a).
The arrow indicates a jump in the $x$-direction by a pulse input and 
the cross denotes a fixed point, $(x,y)=(\frac{I}{101},\frac{10I}{101})$.
The dashed line is the threshold.
Here, both neurons follow the same orbit, so the 
two orbits overlap in the figure.
In the following, we theoretically examine in what parameter regions 
anti-phase states exist stably.

\subsection{Existence of anti-phase states}
Before we discuss the stability of anti-phase states,
we derived a theory about the region of existence of anti-phase states.

We can obtain the solution orbits of anti-phase states analytically.
Since the resonate-and-fire model is linear except for the moment of firing, 
we can integrate it piecewise. 
For ease of explanation, consider the imaginary plane and define
\begin{equation}
z\equiv x+ \mathrm{i} y.
\end{equation}
Then, the resonate-and-fire model can be written as,
\begin{equation}
\frac{d}{dt}(z-z_*)=(-1+10\mathrm{i})z+I=(-1+10\mathrm{i})(z-z_*),
\end{equation}
where,
\begin{equation}
z_* = \frac{I}{101}+\frac{10I}{101}\mathrm{i}.
\end{equation}
We can integrate it easily to,
\begin{equation}
z(t) = z_* + (z_0 - z_* ) e^{ (-1+10\mathrm{i}) t },
\label{eq.1}
\end{equation}
where $z_0$ denotes an initial condition at $t=0$.
For simplicity, let us consider an orbit that is reset at $t=0$.
The reset value is $(x,y)=(0,-1)$. 
Therefore, the initial condition is $z_0=-i$ in the complex plane.

We assume the orbit receives a pulse input and jumps 
in the $x$-direction at $t=T$.
Let $t=T$ in Eq.~(\ref{eq.1}) and add $K$.
Then, the neuronal state just after the pulse input becomes
\begin{equation}
z(T+0) = z_* + (-\mathrm{i} - z_* ) e^{ (-1+10\mathrm{i}) T } + K.
\label{z(T+0)}
\end{equation}
Then, the orbit evolves to $t=T+T'$.
The orbit at $t=T+T'$ is written using Eq.~(\ref{z(T+0)}) as,
\begin{equation}
z(T+T') = z_* + (z(T+0) - z_* ) e^{ (-1+10\mathrm{i}) T' }.
\end{equation}

The threshold line is $y=1$ and the imaginary part of $z$ is $y$.
Thus, the condition for firing time is,
\begin{equation}
y(T+T')=\mbox{Im}(z(T+T'))=1.
\label{y}
\end{equation}
The orbit with $T'=T$ matches a situation where two neurons fire 
alternately at regular intervals.
Thus, the condition that the orbit is an anti-phase state is,
\begin{equation}
y(2T)=1.
\label{condition}
\end{equation}
Next, to set a limit to the $T$ range satisfying $y(2T)=1$ theoretically,
we used the following theorem.
\begin{theorem}

In anti-phase states, a neuron must fire within 
360-degree rotation around the fixed point after a pulse input.
\end{theorem}
\begin{proof}
After 360-degree rotation, the orbit returns to the initial angle around 
the fixed point. 
However, the radius is smaller than the initial one.
Then, the orbit that does not exceed the threshold 
within 360-degree rotation after a pulse input does not exceed forever.
Thus, there is no anti-phase state where neurons do not fire within 360-degree 
rotation after a pulse input. \qed
\end{proof}

This theorem restricts the $T$ range in which to search 
to a finite region.
The restricted region is $0<T<\frac{2\pi}{10}$ because 
the angular velocity around the fixed point is always $10$ and 
it takes $\frac{2\pi}{10}$ to rotate by 360 degrees.

Thus, we should find a $T$ value satisfying $y(2T)=1$ 
in $0<T<\frac{2\pi}{10}$, where the orbit must reach the threshold at 
time $2T$ for the first time after being reset.
The condition that the orbit exceeds the threshold at time $2T$ for the first 
time after being reset can be mathematically represented as 
$y(t)<1$ in $0<t<2T$.

\subsection{Stability of anti-phase states}

In this section, we derive a theory for the stability of anti-phase states.

\begin{figure}[t]
\includegraphics[width=70mm]{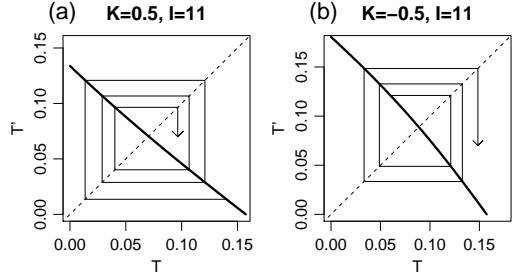}
\caption{\label{spider.ps}
(a)Bold line represents return map with $K=0.5$ and $I=11$.
Dashed line is diagonal. Thin line with arrow is 
construction to obtain firing times. 
$T$ approaches to fixed point by iteration of map.
Anti-phase state is stable.
(b)Return map with $K=-0.5$ and $I=11$. 
$T$ moves away from fixed point by iteration of map.
Anti-phase state is unstable.}
\end{figure}

$T'$ satisfying $y(T+T')=1$ (Eq.~(\ref{y})) can be obtained 
as a function of $T$.
Here $T'$ denotes the interval between pulse input and firing and 
$T$ denotes the interval between reset and pulse input.
We refer to the map that maps T to T' as a return map.
Fig.~\ref{spider.ps}(a) plots the return map with $K=0.5$ and $I=11$.

In anti-phase states, the interval between pulse input and firing 
for one neuron equals the interval between reset and pulse input 
for another neuron.
Thus, we can obtain the firing times of both neurons by using 
the return map iteratively.
The thin line with the arrow in Fig.~\ref{spider.ps}(a) demonstrates
how to iterate the return map.

The intersection between the return map and diagonal line represents
an anti-phase state and is a fixed point on the return map.
For the return map in Fig.~\ref{spider.ps}(a), the orbit finally arrives 
at the fixed point starting from any initial value 
and the anti-phase state is globally stable. 
However, Fig.~\ref{spider.ps}(b) demonstrates that the anti-phase 
state with $K=-0.5$ and $I=11$ is unstable.

Although we can determine global stability by drawing return maps,
we have to redraw them when we change the values of $K$ and $I$.
Therefore, it is difficult to monitor global stability 
over the whole range of $K$ and $I$.
In the following, we limit our focus to local stability and 
show that the condition of neutral stability is analytically obtained 
over the whole range of $K$ and $I$.
Local linear stability is determined as described below.
We calculate the slope of the intersection between the return map and the 
diagonal line ($T'=T$).
An anti-phase state is stable, 
if the absolute value of the slope is smaller than 1.
Otherwise, it is unstable.
When the slope is -1, an anti-phase state destabilizes 
through the period doubling bifurcation
\cite{guckenheimer,kuznetsov,hoppensteadt,wiggins}.
When the slope is 1, a pair of anti-phase states disappears 
through saddle-node bifurcation.

We already derived the theory to determine the region of existence of 
anti-phase states in the previous section.
Here, we will derive a condition for the neutral stability 
of anti-phase states.

\begin{figure}[t]
\includegraphics[width=70mm]{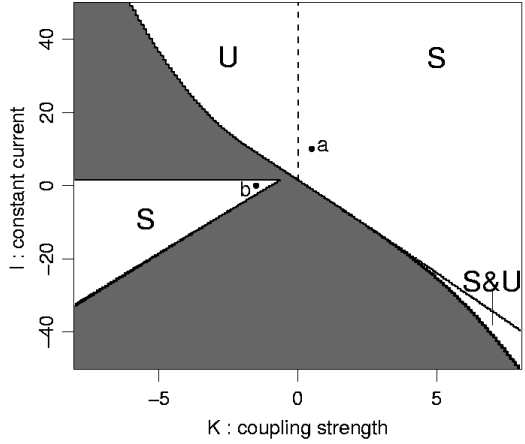}
\caption{\label{existence.ps}
Existence and stability of anti-phase states.
In regions marked ``S'' or ``U'', 
anti-phase state is stable or unstable.
In region marked ``S\&U'', there are two types of anti-phase states.
One of these is stable and the other is not.
Dashed line is neutral stability ($K=0$).
There is no anti-phase state in dark region.}
\end{figure}

We define $f(T,T')$ as a function of $T$ and $T'$,
\begin{equation}
f(T,T') \equiv y(T+T')=\mbox{Im}(z(T+T')),
\end{equation}
where $y(T+T')$ is defined in Eq.~(\ref{y}).
An anti-phase state is neutrally stable if the slope of its return map 
at the fixed point is -1, i.e.,
\begin{equation}
\frac{dT'}{dT}=-\frac{(\frac{df}{dT})}{(\frac{df}{dT'})}=-1,
\label{m}
\end{equation}
where the derivative is evaluated at the fixed point.
After additional calculations, the condition for 
neutral stability (Eq.~(\ref{m})) becomes,
\begin{equation}
K(\tan(10T')-10)=0.
\label{simple1}
\end{equation}
Note that Eq.~(\ref{simple1}) only depends on the sign of $K$ and $T'(=T)$,
and can be rewritten as,
\begin{eqnarray}
K&=&0, \\
\label{neutral}
T&=&\mbox{Arctan}(10)/10=0.1471128, \rm{and} \\
T&=&(\mbox{Arctan}(10)+\pi)/10=0.461272,
\end{eqnarray}
where capitalized $\mbox{Arctan}$ denotes the principal value.
An anti-phase state is neutrally stable,
if any of the above three conditions are satisfied.
Although $\mbox{arctan}$ is a multivalued function,
the theorem in the previous section restricts the $T$ range 
to $0<T<\frac{2\pi}{10}$ and we only need to consider two values of $T$.

We can rewrite the three conditions in terms of $K$ and $I$ 
by setting $T=T'=0.1471128$ (or $0.461272$) in Eq.~(\ref{y}).
As a result, the three conditions for neutral stability become,
\begin{eqnarray}
\label{neutral1}
K &=& 0, \\
\label{neutral2}
I &=& -5.056553 K + 1.587449, \rm{and} \\
\label{neutral3}
I &=&  4.58563 K + 4.461462.
\end{eqnarray}

\section{\label{sec4}results}

We calculated the phase diagram for existence and stability of 
anti-phase states in the $KI$ plane based on our theory.
The results are plotted in Fig.~\ref{existence.ps}.
The anti-phase state is stable in the region marked ``S'' and unstable in 
that marked ``U''.
In the region marked ``S\&U'', there are two types of anti-phase states.
One of these is stable and the other is not.
The dashed line represents neutral stability (Eq.~(\ref{neutral1})).
There are no anti-phase states in the dark region.

To verify the theoretically obtained phase diagram in Fig.~\ref{existence.ps}, 
we did numerical simulation using the Runge-Kutta method
for numerous parameter values.
We assessed an anti-phase state as stable if the orbit starting from it 
remained as it was.
The results of numerical simulation coincided with those of our theory.

\begin{figure}[t]
\includegraphics[width=70mm]{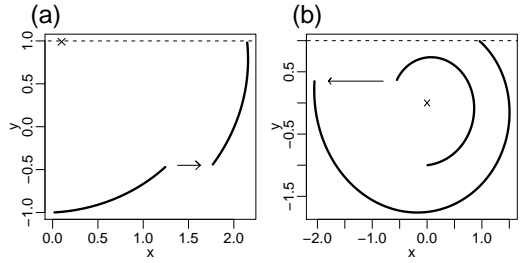}
\caption{\label{snap.ps}Orbits of anti-phase states at 
points marked ``a'' and ``b'' in Fig.~\ref{existence.ps}.
Arrow indicates jump in $x$-direction by pulse input and 
cross denotes fixed point.
Dashed line is threshold ($y=1$).
(a)Orbit with $K=0.5$ and $I=10$.
(b)Orbit with $K=-1.5$ and $I=0$.}
\end{figure}

There are two distinct stable regions in Fig.~\ref{existence.ps}.
To determine the properties of these regions, 
we checked the orbits of anti-phase states at points marked 
``a'' and ``b'' in Fig.~\ref{existence.ps}.
The results are in Fig.~\ref{snap.ps}(a) and (b).

Fig.~\ref{snap.ps}(a) plots the orbit of an anti-phase state with 
$K=0.5$ and $I=10$.
The main characteristics of the orbit are that $K>0$ and period $T$ is 
relatively short.
Since a pulse input just accelerates firing time,
the orbit is not unique to coupled resonate-and-fire models and 
can also be seen in coupled integrate-and-fire models.
The arrow indicates a jump in the $x$-direction by a pulse input and 
the cross denote the fixed point. 
The dashed line is the threshold ($y=1$).

Fig.~\ref{snap.ps}(b) plots the orbit of the anti-phase state with 
$K=-1.5$ and $I=0$.
The main characteristics of the orbit are that $K<0$ and period $T$ is 
relatively long.
The orbit jumps in the negative direction by a pulse input and 
rebounds to fire as if it were a spring.
This is unique to coupled resonate-and-fire models and 
cannot be seen in coupled integrate-and-fire models.
In the region with $I>I_c^1=1.56$, there is no anti-phase state 
such as this, because a neuron spontaneously fires without pulse inputs there.

\begin{figure}[t]
\includegraphics[width=70mm]{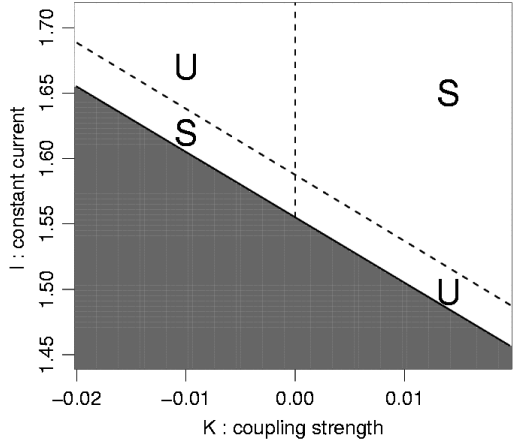}
\caption{\label{phase-mag.ps}
Magnification of Fig.~\ref{existence.ps} around $(K,I)=(0,1.56)$.
In regions marked ``S'' or ``U'', 
anti-phase state is theoretically stable or unstable.
Two dashed lines represent neutral stability 
(Eq.~(\ref{neutral1}) and Eq.~(\ref{neutral2})).
There is no anti-phase state in dark region.
Stability changes near boundary of region of existence 
and on line $K=0$.}
\end{figure}

Fig.~\ref{phase-mag.ps} is the magnification of Fig.~\ref{existence.ps} 
around $(K,I)=(0,1.56)$.
The figure illustrates that stability changes near the boundary of the 
region of existence and on line $K=0$.
In the region marked ``S'', the anti-phase state is stable and 
unstable in ``U''.
The two dashed lines represent neutral stability 
(Eq.~(\ref{neutral1}) and Eq.~(\ref{neutral2})).
There is no anti-phase state in the dark region.

\begin{figure}[t]
\includegraphics[width=70mm]{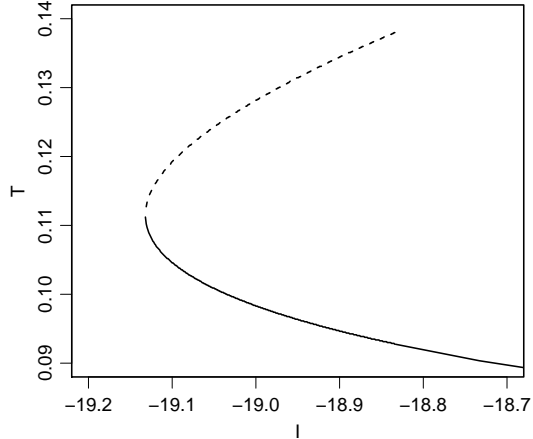}
\caption{\label{saddle.ps}
Period of anti-phase states at $K=4$ as function of $I$.
Two states coexist in region.
Solid line and dashed line denote stable state and 
unstable state, respectively.
At $I=-19.13$, two anti-phase states are created pairwise through 
saddle-node bifurcation.
In $-19.13<I<-18.83$, two states coexist.
At $I=-18.83$, unstable state disappears due to 
effect of threshold. 
In $I>-18.83$, only stable state exists.}
\end{figure}

Two anti-phase states coexist in the region marked ``S\&U'' in 
Fig.~\ref{existence.ps}.
To examine this region, we computed the period of anti-phase states 
at $K=4$ as a function of $I$.
The results are plotted in Fig.~\ref{saddle.ps}.
The solid line denotes a stable state and the dashed line denotes  
an unstable one.
At $I=-19.13$, two anti-phase states are created pairwise through 
the saddle-node bifurcation
\cite{guckenheimer,kuznetsov,hoppensteadt,wiggins}.
In $-19.13<I<-18.83$, the two states coexist.
At $I=-18.83$, the unstable state disappears due to the 
effect of the threshold. 
In $I>-18.83$, only the stable state exists.

\begin{figure}[t]
\includegraphics[width=70mm]{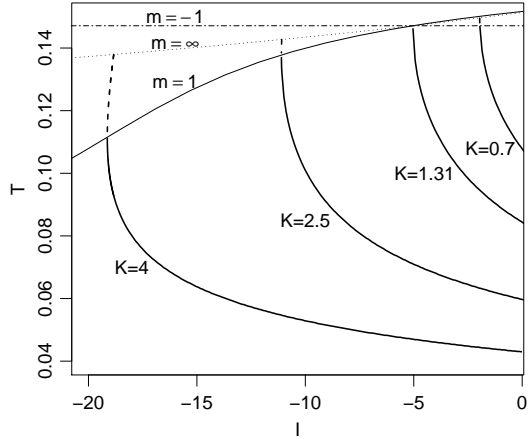}
\caption{\label{stability.ps}
Thick lines are periods of anti-phase states 
as functions of $I$ at different values of $K$ ($K=0.7,1.31,2.5,4$).
Solid line and dashed line denote stable state and 
unstable state, respectively.
Three thin lines are $m=-1,\infty$, and 1,
where $m$ is defined as left hand side of Eq.~(\ref{m}), i.e.,
$m=\frac{dT'}{dT}$.
At $m=1$, two anti-phase states are created pairwise through 
saddle-node bifurcation of return map.
At $m=-1$, anti-phase state destabilizes through 
period doubling bifurcation.
Condition that $m=-1$ is Eq.~(\ref{neutral}).
At $m=\infty$, anti-phase state disappears 
because orbit is tangential to threshold.
In $K>K_c(=1.31)$, two anti-phase states 
with different periods can coexist for some $I$ ranges.
One of these is stable and the other is not.
In $K<K_c$, anti-phase state destabilizes 
crossing line of $m=-1$.}
\end{figure}

Fig.~\ref{stability.ps} plots the period of anti-phase 
states as a function of $I$ at different values of $K$.
The thick solid line denotes a stable state and the thick dashed line denotes 
an unstable one.
The three thin lines are $m=-1,\infty$, and $1$.
Here $m$ is defined as the left hand side of Eq.~(\ref{m}), i.e.,
$m=\frac{dT'}{dT}$.
At $m=1$, two anti-phase states are created pairwise through 
the saddle-node bifurcation of the return map.
At $m=-1$, the anti-phase state destabilizes through 
the period doubling bifurcation.
Here, the condition that $m=-1$ becomes Eq.~(\ref{neutral}).
At $m=\infty$, the anti-phase state disappears 
because the orbit is at a tangent to the threshold.
In $K>K_c(=1.31)$, two anti-phase states 
with different periods can coexist for some $I$ ranges.
One of these is stable and the other is not.
In $K<K_c$, the anti-phase state destabilizes 
crossing the line of $m=-1$.
This is consistent with Fig.~\ref{phase-mag.ps}.

In $K<0$ and $I<I_c^1(=1.56)$, 
there is also a region where two anti-phase states coexist and there is
an unstable region near the boundary of the region of existence,
as in $K>0$.
However, these regions are so small that we cannot see them 
in Fig.~\ref{existence.ps}.

\section{summary and discussion}

In this paper, we analyzed a system for two pulse-coupled resonate-and-fire 
models.
We found that the system settled to an anti-phase state 
in numerical simulations.
We looked for the existence and evaluated the stability of anti-phase states.
We found an effective method of calculating the region of existence, 
where we set limits for the region theoretically 
and then utilized analytically obtained orbits.
We found that stability of anti-phase states could be determined 
by means of a return map of firing times.
The condition for neutral stability turned out to be unexpectedly 
simple (Eq.~(\ref{simple1})).
Based on our theory, we calculated a phase diagram for the existence and 
local stability of anti-phase states.

Stability changed near the boundary of the region of 
existence or at $K=0$.
We also determined stability by direct numerical simulations, and 
the results coincided with that of our theory.
The phase diagram revealed that there were two types of anti-phase states.
An anti-phase state with a longer period is unique to coupled 
resonate-and-fire models and cannot be seen in coupled 
integrate-and-fire models.

We modeled a spike as a delta function.
The case where a spike is modeled as an alpha function has been 
studied in \cite{garbo}.
However, stability condition was not obtained explicitly, 
and a phase diagram for existence of anti-phase states was not 
calculated as a function of coupling strength.
Therefore, it is difficult to compare their results with ours.

In this paper, we focused on anti-phase states.
What about in-phase states?
If two neurons start under the same initial conditions, they must continue to 
follow the same orbit.
Then, in-phase states exist at $I>I_C^1 (=1.56)$ 
where a neuron can fire spontaneously.
However, in-phase states are unstable against perturbations.
This can be explained as follows.
Let the orbits of two neurons differ infinitesimally.
When the neurons fire in succession, the leading neuron receives a pulse 
input from the following neuron just after being reset ,
while the following neuron receives a pulse input from the leading neuron 
just before firing.
The positions where they receive pulse inputs differ considerably.
The effects of pulse inputs on orbits are so different that the 
difference between orbits becomes finite.
Thus in-phase states are always unstable.

\begin{acknowledgments}
We are grateful to S. Shinomoto and Y. Kuramoto for valuable discussions and 
suggestions of this work.
This work was partially supported by Grant-in-Aid for Scientific Research 
on Priority Areas No. 14084212.
\end{acknowledgments}

\bibliography{pre}% Produces the bibliography via BibTeX.

\end{document}